\documentclass{elsart}
 
\usepackage{epsfig,subfigure,latexsym,amsmath}
 
\begin{document}
 
\input epsf
 
\begin{frontmatter}

\title{A nucleonic NJL  model for finite nuclei: dynamic mass generation and ground-state observables}
\author[lanl]{T. J. B{\"u}rvenich}
\author[lanl]{and D. G. Madland}
\address[lanl]{Theoretical Division, Los Alamos National Laboratory, Los Alamos, New Mexico 87545}

\begin{abstract}
We test the compatibility of chiral symmetry, dynamic mass generation of the nucleon
due to spontaneous breaking of chiral symmetry, and the description of finite nuclear systems by employing
an NJL model understood as a chiral invariant effective theory for nucleons. We apply the model
to nuclear matter as well
as to finite nuclei. In the latter case,
the model is adjusted to nuclear ground-state observables.
We treat the case of a pure chiral theory and the physically more realistic case where a portion of
the nucleon mass (160 MeV) explicitly breaks chiral symmetry.
The best version of this current model is found to deliver
reasonably good results simultaneously
for both finite nuclei and the nucleon mass,
which supports our motivation of probing a link between
low-momentum QCD and the nuclear many-body problem.
However, the observables calculated for finite nuclei are not as good as those
coming from existing relativistic mean field models without explicit
chiral symmetry.
\end{abstract}
 
\end{frontmatter}
 
\section{Introduction}
 
QCD is believed to be the fundamental theory of the strong interaction. It has non-perturbative
behaviour at low energies. One of its important non-perturbative phenomena is the nontrivial
vacuum structure:  the vacuum
is populated by scalar quark-antiquark
pairs leading to a finite expectation value of $\langle \bar{q}q \rangle$,
the chiral or quark condensate.
The Nambu-Jona-Lasinio (NJL) model \cite{Nam61a,Nam61b,Vog91,Kle92}, originally invented for nucleonic,
but mostly used for quark degrees of freedom, schematically implements this rearrangement of the vacuum.
The quark mass is generated by spontaneous symmetry breaking due to the
negative energy states of the Dirac sea which contribute to the scalar density. In the context of
nucleonic degrees of freedom, the major part of the nucleon mass is generated by the chiral condensate,
which is modeled (see below for a more specific explanation) by $\langle\bar{N}N\rangle$, where $N$ denotes nucleon states. A smaller part of the
nucleon mass emerges from an explicit chiral symmetry breaking contribution.
 
A link between hadronic effective field theories for nuclear matter and finite nuclei and
QCD can be established via the symmetries of (massless) QCD: predominantly Lorentz invariance
and chiral symmetry. There exist models employing a nonlinear
realization of chiral symmetry that have proven to be successful \cite{Fur97,Sch02}.
Successfull models for nuclei not considering explicit chiral invariance are relativistic
mean-field (RMF) models, either the Walecka-type (RMF-FR) with explicit finite range meson
exchange \cite{Ser86,Rei89}  or the point-coupling
type (RMF-PC) with contact interactions and derivative terms to simulate finite range
\cite{NHM92,Buer02,Man88}.
Both of these deliver accurate descriptions of
finite nuclei observables such as binding energies, densities, radii, surface thicknesses, and
ground-state deformations. They can be considered compatible with  chiral symmetry
in the sense that power counting (NDA - naive dimensional analysis) based upon chiral symmetry
works. This is more transparent in the point-coupling variant of these models.
Thus, nuclei are compatible with QCD scales \cite{Fri96,Buer02,Rus97}.
 
By describing nuclear matter with NJL models one employs effectively (and schematically) three features
of QCD,  namely, chiral symmetry,
the mass generation due to its explicit and spontaneous breaking, as well as the binding of nuclear
matter (i.e. the nucleon in-medium self-energy) resulting from the change of the condensate.
Nonlinear NJL models for nucleon degrees of freedom have been applied to nuclear matter
\cite{Koch1,Bent01,Mos02,Mis03} and,
in a simple version, to
finite nuclei \cite{Koch2}.
Nucleonic NJL models provide an attempt to incorporate chiral symmetry and its dynamical breaking within the framework of
relativistic mean-field models.
The chiral condensate is modeled in terms of nucleonic
degrees of
freedom, $\langle \bar{q}q\rangle \rightarrow \langle\bar{N}N \rangle$.
Some comments on this relation are in order. It has been shown by several authors \cite{Fur97,Jar91,Ble87,Coh92,Coh89} that expressing the
quark Dirac sea by the nucleon Dirac sea is inapproriate. Strictly speaking, the nucleonic degrees of
freedom, which are well suited for the valence particles in the spirit of effective field theory, are the wrong ones for describing the chiral
condensate.
Replacing $\langle\bar{q}q\rangle$ by
$\langle\bar{N}N \rangle$ is not compatible with the large $N_c$ limit of QCD. Furthermore,
the composite nature of the nucleon suppresses corresponding vacuum loops.
In our case, we do not think of a literal replacement. The sole model property needed stemming from
$\langle\bar{N}N \rangle$ is a large (negative) contribution to the scalar density
(the scalar mean field) which in our best-fit model contributes to about 80 \% of the nucleon mass. This usage of the
nucleon Dirac sea in finite systems is put to the test.

Based on our experience with RMF-PC models \cite{Fri96,Buer02}, we would like to test an extended and
physically more realistic
version for finite nuclei, including important ingredients such as derivative terms, the Coulomb force,
and the center of mass
correction. Nuclei provide a more stringent test of this model than infinite nuclear matter does since
the shell structure and density distribution are quite sensitive to the size and the density dependence
of the mean-field potentials.
Our basis for comparison will be the RMF-PC force PC-F1 determined in Ref. \cite{Buer02} which
respresents the accuracy that can be obtained in modern mean-field approaches.
We call the model used here, to
distinguish it from the NJL model for quark degrees of freedom, the nucleon NJL (NNJL)
model (we correspondingly denote that for quarks by qNJL). Note that a model for
the deuteron has also been denoted as NNJL \cite{Iva00}.
 
An NJL approach for nuclei resembles the RMF-PC model in the Hartree approximation which
we also use here, since all densities and currents containing
$\gamma_5$ matrices vanish in ground states of even-even nuclei. The most prominent
difference between the two is that RMF models contain an explicit
mass term and are treated in the no-sea approximation, whereas the mass is dynamically
generated within NJL
models. Another difference is that in contrast to the linear RMF-PC
model, the linear NJL model cannot describe saturation of nuclear matter
(if reproducing the vacuum nucleon mass at the same time). Saturation can only be obtained
at the price of a small or vanishing nucleon mass \cite{Mishus97}. Nonlinear NJL models, however,
are capable of simultaneously describing saturation and the vacuum nucleon mass, as
demonstrated by Koch et al. \cite{Koch1}, as well as
the chiral phase transition.
Another important difference is that imposing chiral symmetry in
the NJL model lagrangian restricts allowed terms. The building blocks of the NJL model that in the Hartree
mean-field approximation reduce to isoscalar terms are the chiral invariant
combinations $\Big[(\bar{\psi}\psi)^2 - (\bar{\psi}\gamma_5\vec{\tau}\psi)^2\Big]$ and
$\Big[(\bar{\psi}\gamma_\mu\psi)^2 + (\bar{\psi}\gamma_5\vec{\tau}\gamma_\mu\psi)^2\Big]$.
This constrains the density dependence of the effective mass (and hence the scalar potential).
 
With our approach, which to our knowledge is the first realistic NJL model for light and heavy finite nuclear
systems in the mean-field
approximation, we wish to probe a linkage between low-momentum QCD and the properties of
finite nuclei by simultaneously employing a chirally symmetric interaction lagrangian and mass
generation by spontaneous breaking of chiral symmetry.
We do not intend to create an effective field theory for nuclei that is better
than the already established models, like the relativistic mean-field model and its chiral
versions.
 
\section{The model}
 
The starting point of our investigations is based upon the nonlinear chiral invariant model
of Koch et al. \cite{Koch1,Koch2}, which we formulate in its $SU(2)_V\otimes SU(2)_{A}$ form
(where $m_{0}=0$):
\begin{equation}
{\mathcal{L}}_{K} = {\mathcal{L}}_{free} + {\mathcal{L}}_{4f} + {\mathcal{L}}_{8f}
\end{equation}
\begin{eqnarray}
{\mathcal{L}}_{free} & = &  \bar{\psi}(i\gamma_\mu\partial^\mu - m_{0}) \psi \nonumber \\
{\mathcal{L}}_{4f} & = & g_1 \Big[(\bar{\psi}\psi)^2 - (\bar{\psi}\gamma_5\vec{\tau}\psi)^2\Big]
- g_2 \Big[(\bar{\psi}\gamma_\mu\psi)^2 + (\bar{\psi}\gamma_5\vec{\tau}\gamma_\mu\psi)^2\Big] \nonumber
\\
{\mathcal{L}}_{8f} & = & g_3 \Big( \Big[(\bar{\psi}\psi)^2 - (\bar{\psi}\gamma_5\vec{\tau}\psi)^2\Big]
\times \Big[(\bar{\psi}\gamma_\mu\psi)^2 + (\bar{\psi}\gamma_5\vec{\tau}\gamma_\mu\psi)^2\Big] \Big)
\end{eqnarray}
The cross-density term associated with the coupling constant $g_3$ is usually missing in RMF models
since
saturation is achieved already in the linear version. Not so in the NNJL model: here the cross
term is necessary to achieve saturation as well as a realistic vacuum mass at the same time
(for a discussion of
the problem of matter stability within the NJL model see \cite{Bub96}).
In the original version from Koch et al., where $m_{0}$ is set to zero, the model has no
explicit symmetry breaking terms. In our analysis, we consider the realistic case with an
explicit symmetry breaking term ($m_0 > 0$).
For comparisons to
the model of Koch et al. in nuclear matter  we also consider the limit of massless nucleons.

We extend ${\mathcal{L}}_K$ for nuclear matter with the following terms:
\begin{eqnarray}
{\mathcal{L}}_{ext1} & = & g_4 \Big[(\bar{\psi}\psi)^2 - (\bar{\psi}\gamma_5\vec{\tau}\psi)^2\Big]^2
- g_5 \Big[(\bar{\psi}\gamma_\mu\psi)^2 + (\bar{\psi}\gamma_5\vec{\tau}\gamma_\mu\psi)^2\Big]^2
\end{eqnarray}
This is motivated by the fact that the effective mass in the model of Koch et al. \cite{Koch1} is too
large ($m^*/m_{vac} = 0.92$) and must be smaller to generate spin-orbit splittings in nuclei
of the right magnitude.
In contrast to the RMF-PC approach, the parameters $g_1$, $g_3$, and $g_4$ not only
govern the scalar potential felt by the nucleons, but simultaneously contribute
to the nucleon mass at zero baryon density.
This is an important additional constraint which does not appear in the case of RMF-PC.
 
For finite nuclei, in a similar fashion, an isovector-vector term is added (having
the quantum numbers like the $\rho$-meson) as well as derivative terms. The
derivatives impose no problem since chiral symmetry is a global symmetry. They are an important and necessary
ingredient in RMF-PC
approaches and similarly in the Skyrme-Hartree-Fock (SHF) approach \cite{Que78} to simulate
finite range and to characterize the surface
region of the nucleus.
We also add the Coulomb
field.
The extension reads:
\begin{eqnarray}
{\mathcal{L}}_{ext2} & = & g_6 \Big[ \partial_\mu (\bar{\psi}\psi)\partial^\mu(\bar{\psi}\psi) -
 \partial_\mu
(\bar{\psi}\gamma_5\vec{\tau}\psi)\cdot\partial^\mu(\bar{\psi}\gamma_5\vec{\tau}\psi) \Big]\nonumber \\
&-& g_7 \Big[ \partial_\mu (\bar{\psi}\gamma_\nu\psi)\partial^\mu(\bar{\psi}\gamma^\nu\psi) +
\partial_\mu
(\bar{\psi}\gamma_5\vec{\tau}\gamma_\nu\psi)\cdot\partial^\mu(\bar{\psi}\gamma_5\vec{\tau}\gamma^\nu\psi) \Big]\nonumber \\
& - & g_8  \Big[(\bar{\psi}\gamma_\mu\vec{\tau}\psi)\cdot (\bar{\psi}\gamma^\mu\vec{\tau}\psi) +
(\bar{\psi}\gamma_\mu\gamma_5\psi)(\bar{\psi}\gamma^\mu\gamma_5\psi)\Big] \nonumber \\
&-& g_9  \Big[\partial_\nu(\bar{\psi}\gamma_\mu\vec{\tau}\psi)\cdot \partial^\nu(\bar{\psi}
\gamma^\mu\vec{\tau}\psi) +
\partial_\nu(\bar{\psi}\gamma_\mu\gamma_5\psi)\partial^\nu(\bar{\psi}\gamma^\mu\gamma_5\psi)\Big]
 \nonumber \\
& - &  e A_\mu \bar{\psi}\Big(\frac{1-\tau_3}{2}\Big)\gamma^\mu \psi - \frac{1}{4}F_{\mu\nu}F^{\mu\nu}
\end{eqnarray}
When applying to finite nuclei, these terms are necessary for RMF-PC models and constitute a minimal
extension as shown
in \cite{NHM92,Buer02}.
The parameters of the model are the nine coupling constants $g_1$, $g_2$, $g_3$, $g_4$, $g_5$,
$g_6$, $g_7$, $g_8$, $g_9$
as well as the (sharp) cutoff $\it\Lambda_{\rm c}$ (see below). We regard this approach as a
phenomenological model in which
all coupling constants are determined by least-squares adjustments to measured nuclear
ground-state observables. We do not take into acount
low-energy theorems. One could, in principle, derive a generalized Gell-Mann-Oakes-Renner (GOR) relation
as can be done in linear $\sigma-\omega$ models \cite{Koch97} by assuming that the vacuum expectation value of the 
explicitly symmetry breaking term is equal to the corresponding term in low-momentum QCD, see also \cite{Mishus97,Mis03}. The Goldstone boson in our model, however, is
an antinucleon-nucleon state and thus can hardly be identified with the pion. We feel that these questions
(at least at this stage) ask too much of this phenomenologically based approach.
 
The NNJL model resembles the phenomenological linear $\sigma-\omega$ model \cite{Koch97,Pok00,Fur93,Fur96} when
replacing the fields with their lowest order approximation, namely $\sigma \rightarrow \bar{\psi}\psi$ and
$\vec{\pi} \rightarrow \bar{\psi}\gamma_5 \vec{\tau} \psi$
(in finite systems additional derivative terms become necessary).
The spontaneous breaking of chiral symmetry in linear $\sigma-\omega$ models is usually realized by assigning a finite
expectation value to the $\sigma$ field due to a mexican-hat type potential and then shifting the field.
In our case, the NJL mechanism leads to a finite vacuum expectation value of the scalar density.
Both models share the linear realization of chiral symmetry.
 
\section{Nuclear matter}
Our nuclear matter model is
\begin{equation}
{\mathcal{L}}_{nm} = {\mathcal{L}}_{K} + {\mathcal{L}}_{ext1}
\end{equation}
containing 5 coupling constants.
Derivative terms as well as isovector terms do not contribute in infinite
symmetric nuclear matter, and nucleons are momentum eigenstates (plane waves).
The effective mass is given by
\begin{equation}
m^* = m_{0} -2 (g_1 + g_3 \rho_V^2 +2 g_4 \rho_S^2) \rho_S
\end{equation}
In this equation, $\rho_V$ denotes the vector density and $\rho_S$ denotes the
scalar density, consisting of the valence part and the negative energy states:
\begin{equation}
\rho_S = \frac{\gamma}{(2\pi)^3} \int_0^{k_F} d^3 k \frac{m^*}{\sqrt{k^2 + {m^*}^2}}
 -  \frac{\gamma}{(2\pi)^3} \int_0^{\it \Lambda_{\rm c}} d^3 k \frac{m^*}{\sqrt{k^2 + {m^*}^2}}
\end{equation}
where $\gamma$ is the spin-isospin degeneracy factor ($\gamma = 4$ in symmetric nuclear matter)
and $k_F$ is the Fermi momentum
of the valence (positive energy) nucleons. We can imagine the sharp three-momentum cutoff
$\it\Lambda_{\rm c}$ as separating the {\em active}
negative energy states from the ones that do not contribute to the scalar density.
Stated differently, the effective
interaction is only sensitive to states up to the momentum $\it\Lambda_{\rm c}$. We call this the
{\it no deep-sea approximation}.
 
The binding energy per particle $B/A$ is given by
\begin{equation}
B/A = \Big(T_{00} - T_{00}^{vac}\Big)/\rho_V - m_{vac}
\end{equation}
where $T_{00}$ is the $00$ component of the canonical energy-momentum tensor.
$ T_{00}^{vac}$ denotes the vacuum contribution that can be obtained by setting
the vector density to zero and using the vacuum nucleon mass $m_{vac}$.
We obtain
\begin{eqnarray}
T_{00}  &=& \int_{0}^{k_F} d^3k \sqrt{k^2 + {m^*}^2} -  \int_{0}^{\it \Lambda_{\rm c}} d^3k \sqrt{k^2
+ {m^*}^2} \nonumber \\
    & + & g_1 {\rho_S}^2 + g_2 {\rho_V}^2 + g_3 {\rho_S}^2 {\rho_V}^2 \nonumber \\
    &+& 3 g_4 {\rho_S}^4 + g_5 {\rho_V}^4
\end{eqnarray}
and
\begin{eqnarray}
T_{00}^{vac} & = &  -  \int_{0}^{\it \Lambda_{\rm c}} d^3k \sqrt{k^2
+ {m_{vac}}^2} + g_1 \rho_{Svac}^2 + 3 g_4 \rho_{Svac}^4
\end{eqnarray}
Note that the constant shift of the negative energy states caused by the vector
potential does not contribute since it is becomes zero by the normal
ordering of the baryon number operator. The integrals for the scalar density and the two binding energy
contributions are done analytically.
The energy computed within the NNJL model results from two large
cancellations: one is, as in the RMF approach, the cancellation of large attractive scalar
and repulsive vector fields and their contribution to the total energy, while the other is
the difference of the contribution at finite baryon density and the vacuum contribution
from the active Dirac sea states.
 
From the experience with RMF models we know that one of the attractive
features of this approach is that the spin-orbit splittings of nuclei
are nicely reproduced, courtesy of the large (several hundred MeV) scalar
and vector potentials. Since with ${\mathcal{L}}_K$, the effective mass
is about $m^*/m_{vac} = 0.92$ at saturation \cite{Koch1}, the spin-orbit splittings in finite
nuclei can be expected to be far too
small (we will demonstrate this statement in the next section).
To get closer to a model which also has a satisfactory predictive power for finite nuclei,
especially their shell structure,
several least-squares adjustments were performed to pseudo-observables $\rho_V = 0.17$~fm$^{-3}$~(0.5),
 E/A = -16.5 MeV~(1.0), $m_{vac} = 938.9$~MeV~(1.0), and $m^*/m = 0.6$ ~(2.0)
[the numbers in parentheses indicate the uncertainties in percent entering the $\chi^2$].
The uncertainties are,
of course, rather arbitrary and reflect only the accuracy that is expected of the pseudo-observables.
The baryon density $\rho_V = 0.17$~fm$^{-3}$ has been favoured over $\rho_V = 0.16$~fm$^{-3}$ since
this value has been used in the adjustments of Ref. \cite{Koch1}.
For the adjustment N-1 a compressibility of $K = 240$~MeV, with 5 \% error, has been
included in  $\chi^2$.
 
Our results are summarized in Tables \ref{numa1} and \ref{numa2}.
These adjustments have been performed with various fixed values for $\it\Lambda_{\rm c}$ since it turned out
that when
including it as a free parameter it does not experience significant changes.
Furthermore, an explicit chiral symmetry breaking mass term improved the $\chi^2$ in the adjustments of N-1 and N-2
compared to N-K$^*$.
According to \cite{Ji95}, where the deep-inelastic momentum sum rule and the trace anomaly of the QCD
energy-momentum tensor have been used to separate different portions of the nucleon mass, this mass term should not
be larger than $m_{0} = 160$~MeV. It turned out,
however, that
larger values are more favorable than smaller ones, so the adjustments N-1 and N-2 have $m_{0}$ set to 160 MeV.
This is a first trial value that has also been employed
in Ref. \cite{Mishus97}. A different choice would be the pion-nucleon sigma term ( $\Sigma \approx 50$~MeV) which we
postpone to forthcoming work.
\begin{table}[htb]
\caption{Bulk properties of nuclear matter for the set N-K, the parameter sets obtained in least-squares
adjustments to nuclear matter properties (N-K$^*$, N-1, N-2), and to finite nuclei (N-FN), as well as for the RMF-PC
force PC-F1.}
\label{numa1}
\begin{tabular}{c|ccccc}
    Set &           $m_{vac}$ [MeV] & $\rho_V$~ [fm$^{-3}]$   &  E/A [MeV] &  $m^*/m_{vac}$   &  K [MeV] \\\hline
 N-K  &   942.4          &   0.172                 &  -16.6     &   0.92     &  354     \\
 N-K$^*$    &   935.7          &   0.167                 &  -16.6     &   0.82     &  355     \\\hline
 N-1   &   937.1          &   0.172                 &  -16.3     &   0.66     &  311     \\
 N-2   &   935.0          &   0.178                 &  -16.5     &   0.72     &  349     \\
\hline
 N-FN &  941.2         &   0.155                 & -16.4      &   0.75     &  388    \\ \hline
 PC-F1      &  938.9         &  0.151                  &  -16.2     &   0.61     & 270     \\
\end{tabular}
\end{table}
\begin{table}[htb]
\caption{Coupling constants of the parameter sets discussed in the text.}
\label{numa2}
\begin{tabular}{c|ccccccc}
  Set &            $g_1$ [fm$^2$]    &      $g_2$   [fm$^2$]   &      $g_3$   [fm$^8$]   &      $g_4$   [fm$^8$]     &     $g_5$
    [fm$^8$]  &    $\it\Lambda_{\rm c}$ [fm$^{-1}$] &   $m_{0}$ [MeV] \\\hline
N-K  &     1.16     &      3.40      &       0.92     &      0.0         &     0.0      &    3.25       &     0.0   \\
N-K$^*$  &     1.5916   &      3.1950    &       1.3665   &      0.0773      &     0.0      &    2.75       &  0.0    \\  \hline
N-1  &     3.3064   &      3.9531    &       2.2270   &      0.9910      &    -0.8917   &    2.00       &   160.0 \\
N-2  &     1.7678   &      3.1279    &       1.0908   &      0.4685      &     0.0      &    2.35       &   160.0  \\
\end{tabular}
\end{table}
\begin{figure}[htb]
\epsfig{figure=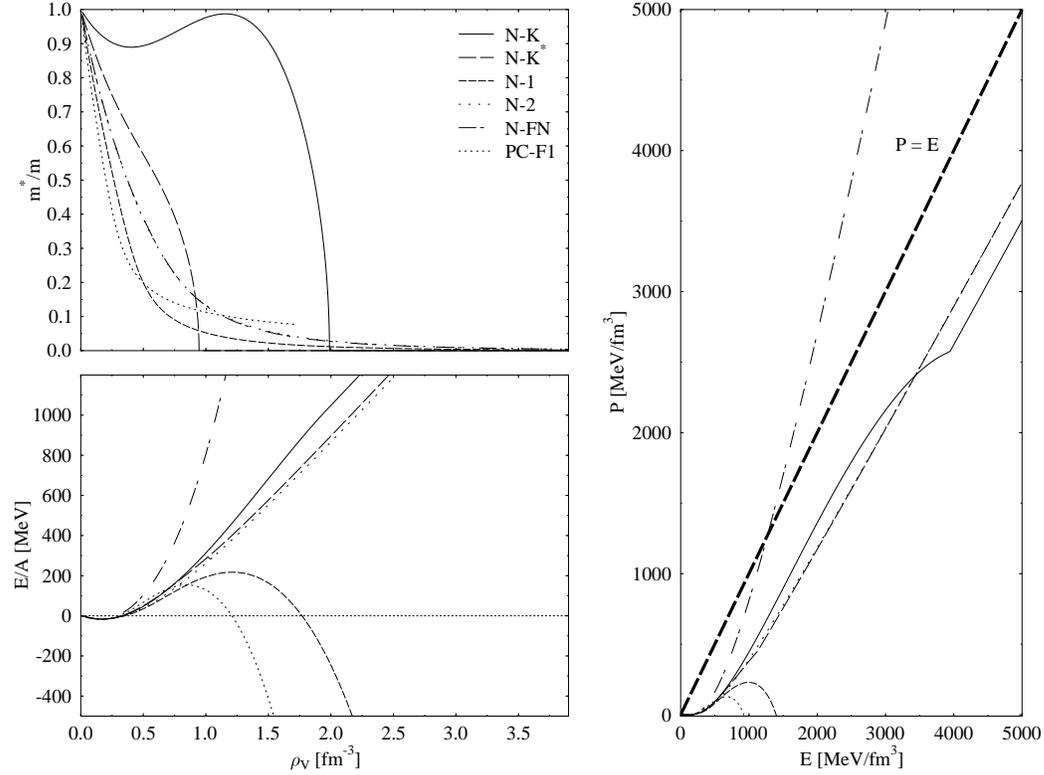,width=14cm}
\caption{Energy per particle (bottom left) and effective mass (top left) as a function of nucleon
density as well as pressure vs. density (right) for the six models discussed in the text. The causal limit
is given by the line P$=$E.}
\label{eos}
\end{figure}
Clearly, an explicit mass term and a cutoff below 2.40 fm$^{-1}$ (N-1 and N-2) achieve an effective mass
which is close (but still
larger) to those determined in RMF models ($m^*$ is larger with $m_0 = 0$, see N-K$^*$). But by comparing N-K$^*$ to N-K we notice
 that the additional terms and the
inclusion of the effective mass in the adjustment procedure drive the model into the right direction already.
 The new
sets N-1, N-2, and N-K$^*$ tend to a smaller vacuum mass, a higher
saturation density and a rather high compressibility compared to the desired values. The energy at saturation appears to be
reproduced by any of the sets of coupling constants. It is curious that the model delivers a high compressibility regardless
of the number and/or values of the parameters. A lower compressibilitiy, K=311 MeV, could only be
achieved
with a very low cutoff value of $\it\Lambda_{\rm c} = 2.0$~fm$^{-1}$ in the set N-1. Overall, N-1 performs best due to the explicit mass term,
the coupling constant $g_5$ and the inclusion of K in the adjustment procedure. The NNJL model appears to favor
cutoff values of approx. 2.0-2.35 fm$^{-1}$. This is still larger than the fermi momentum of the valence nucleons, but
clearly smaller than values used in qNJL models ($\approx 3$~fm$^{-1}$). This empirical fact, which has been found also by other
authors \cite{Mos02,Mis03}, still lacks a clear explanation. The cutoff, however, is of the order of the $\sigma$ meson
mass used in RMF models ($\approx 450-550$~MeV).
 
Figure \ref{eos} shows the effective mass and the energy/particle as functions of the
baryon density as well as pressure versus energy.
Clearly, models with (PC-F1, N-1, N-2) and without (N-K, N-K$^*$) an explicit mass term exhibit a quite different dependence of the
nucleon mass with density. In a pure chirally symmetric theory, there is a second order
phase transition from massive nucleons to massless particles. However, in the physically realistic models with
$m_{0} > 0$, a smooth transition takes place which resembles the behaviour
of the effective mass in RMF models. For a successful model of nuclei in terms
of spectral properties, the effective mass needs i) $m^*/m \approx 0.6$ and ii)
a density dependence that leads to the proper radial
dependence of the spin-orbit potential [which is proportional to the spatial
derivative of $V_S - V_V$, where $V_S$ ($V_V$) is the scalar (vector)
potential]. We note that the RMF-PC force (PC-F1) and the NNJL sets adjusted with respect to nuclear matter pseudo observables
(N-K, N-K$^*$, N-1, N-2) obey causality. Causality is violated above E~$> \approx 1500$~MeV/fm$^3$, however, for the set N-FN which
has been adjusted to nuclear ground-state observables (see the discussion in the next section).
 
We also note from the lower portion of Fig. \ref{eos} that a negative value of $g_5$ (set N-1) softens the equation of state due
to a reduction of the
repulsion by the vector potential. At high densities, however, it leads to an
instability of the equation of state bending downwards again. This is a common feature
of PC-F1 and N-1, and occurs at such high densities ($\rho_V > 1.0 ~{\rm fm}^{-3}$) that it does not limit
the application of these models to ground-state observables of finite nuclei with densities ($\rho_V \approx 0.17 ~{\rm
fm}^{-3}$) that are very near and below the saturation density in the equation of state.
 
\section{Finite nuclei}
 
Our model for finite nuclei is
\begin{equation}
{\mathcal{L}}_{fn} = {\mathcal{L}}_K + {\mathcal{L}}_{ext1} + {\mathcal{L}}_{ext2}
\end{equation}
containing 9 coupling constants.
Finite nuclei are
calculated in coordinate space employing matrix multiplications as derivatives in Fourier space.
The solution with
lowest energy is determined using the damped gradient step method \cite{grad}. Pairing for
the nucleons is not yet invoked.
The negative energy states are, as in \cite{Koch2} and for the previous nuclear matter case,
treated in the LDA
approximation with a spatially dependent effective mass $m^*(\vec{r})$.
The positive energy states are calculated explicitly and selfconsistently. The parameters
containing isovector terms and derivative
terms (both isoscalar and isovector) do not contribute to the Dirac sea states since they
vanish in symmetric
homogenous nuclear matter.
We employ the microscopic center of mass (c.m.) correction
$E_{c.m.} = \frac{<P^2>}{2mA}$ for the positive energy nucleons only. Since the negative energy states are
locally treated as chunks of nuclear matter, their translational invariance renders a zero c.m.
correction.
 
Starting from the parameter set N-2  we performed least-squares adjustments to
ground-state observables
of 7 nuclei and to the vacuum mass $m_{vac}$ of the nucleon simultaneously.
The resultant parameter set we label N-FN. The four observables used were
binding energy E  (uncertainty 0.2 percent), diffraction radius R (uncertainty 0.5 percent),
rms radius r (uncertainty 0.5 percent), and surface thickness $\sigma$  (uncertainty 1.5 percent).
The nuclei and
observables chosen were $^{16}$O (E, R, $\sigma$), $^{40}$Ca (all four), $^{48}$Ca (all four),
$^{88}$Sr (E, R, r), $^{90}$Zr (all four), $^{132}$Sn (E) and $^{208}$Pb (all four). The uncertainty
for the vacuum nucleon mass was chosen to be 1 \% ($\approx 10$~MeV). Using a much larger or smaller uncertainty
affected the adjustment only to a minor extent - the mass, in conjunction with nuclear ground-state
observables, appears to be a robust prediction.
This set of observables (except for the nucleon mass) is a subset of the ones used to extract the
RMF-PC force PC-F1.
\begin{table}[p]
\caption{Observables of the least-squares adjustment procedure: binding energy (E), diffraction
radius (R), rms radius (r) and surface thickness ($\sigma$) (units: [fm] for the radii and surface thickness, [MeV] for the energy). The last line indicates
the total $\chi^2$ with respect to this set of observables. Note that N-K
and PC-F1 have been adjusted earlier to other sets of observables.}
\label{obs}
\begin{tabular}{l|ccccc}
observable           & expt.     &    N-K     &  N-FN    & PC-F1  \\  \hline
 
$m_{vac}$            & 938.9     &  942.4      &   941.2            &  938.9  \\\hline
 
$^{16}$O: E          & -127.6    &   -191.1  &   -127.3         &   -127.7  \\
$^{16}$O: R          & 2.777     &   2.646     &  2.805           &   2.769   \\
$^{16}$O: $\sigma$   & 0.839     &   0.601     &  0.790           &    0.850  \\\hline
 
$^{40}$Ca: E         & -342.1   &   -445.0   &  -345.0         &  -344.9  \\
$^{40}$Ca: R         & 3.845     &   3.693     &   3.862          &   3.839   \\
$^{40}$Ca: r         & 3.478     &   3.173     &   3.418          &   3.453   \\
$^{40}$Ca: $\sigma$  & 0.978     &   0.706     &   0.893          &   0.967        \\\hline
 
$^{48}$Ca: E         & -416.0   &  -530.5    &  -412.1         &   -416.0  \\
$^{48}$Ca: R         & 3.964     &  3.833      &   3.945          &   3.945     \\
$^{48}$Ca: r         & 3.479     &  3.252      &   3.434          &   3.444    \\
$^{48}$Ca: $\sigma$  & 0.881     &  0.712   &   0.867          &    0.885       \\\hline
 
$^{88}$Sr: E         & -768.5   &  -932.1    &   -765.2        &  -769.0    \\
$^{88}$Sr: R         & 4.994     &  4.861      &   4.982          &   5.005    \\
$^{88}$Sr: r         & 4.224     &  3.985      &   4.168          &   4.197        \\\hline
 
$^{90}$Zr: E         & -783.9    &  -941.6   &   -779.9         &   -785.3        \\
$^{90}$Zr: R         & 5.040     &  4.835     &   4.997          &    5.025       \\
$^{90}$Zr: r         & 4.270     &  4.012     &   4.22           &    4.245      \\
$^{90}$Zr: $\sigma$  & 0.957     &  0.796     &   0.939          &    0.943       \\\hline
 
$^{132}$Sn: E        & -1102.9   &  -1376.5  &   -1107.1        &   -1102.8  \\\hline
 
$^{208}$Pb: E        & -1636.4 &   -2004.4  &   -1642.5       &  -1636.9  \\
$^{208}$Pb: R        & 6.806     &   6.775     &   6.742          &   6.808    \\
$^{208}$Pb: r        & 5.504     &   5.330     &   5.438          &  5.501      \\
$^{208}$Pb: $\sigma$ & 0.900      &   0.341     &   0.852          &    0.876        \\\hline\hline
$\chi^2$             &            & 156728      & 174              &     35            \\
\end{tabular}
\end{table}
\clearpage
\begin{figure}[htb]
\epsfig{figure=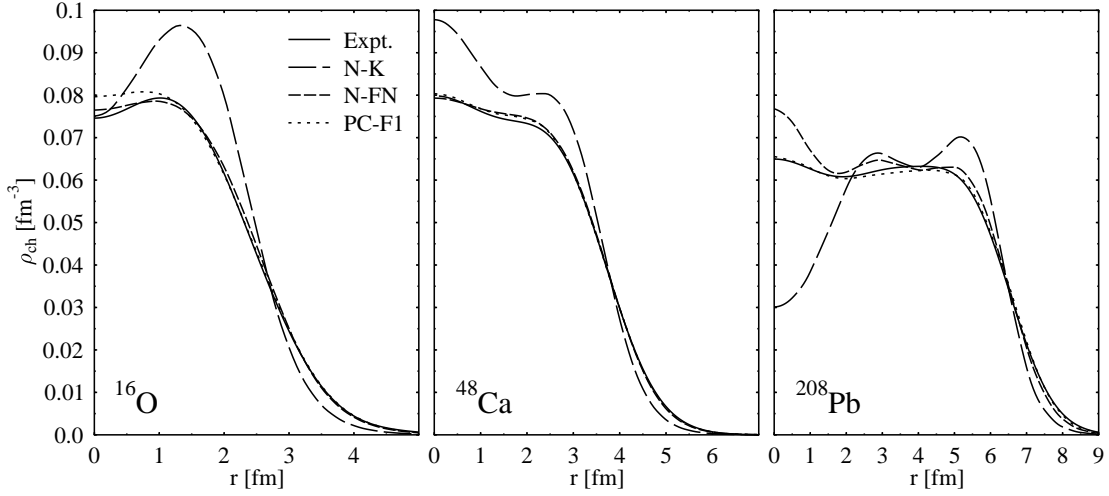,width=15cm}
\caption{Charge densities of the nuclei $^{16}$O (left), $^{48}$Ca (middle) and $^{208}$Pb (right) with the
forces as indicated. The experimental data are from Ref. \protect \cite{Vri87}.}
\label{dens}
\end{figure}
\begin{figure}[htb]
\epsfig{figure=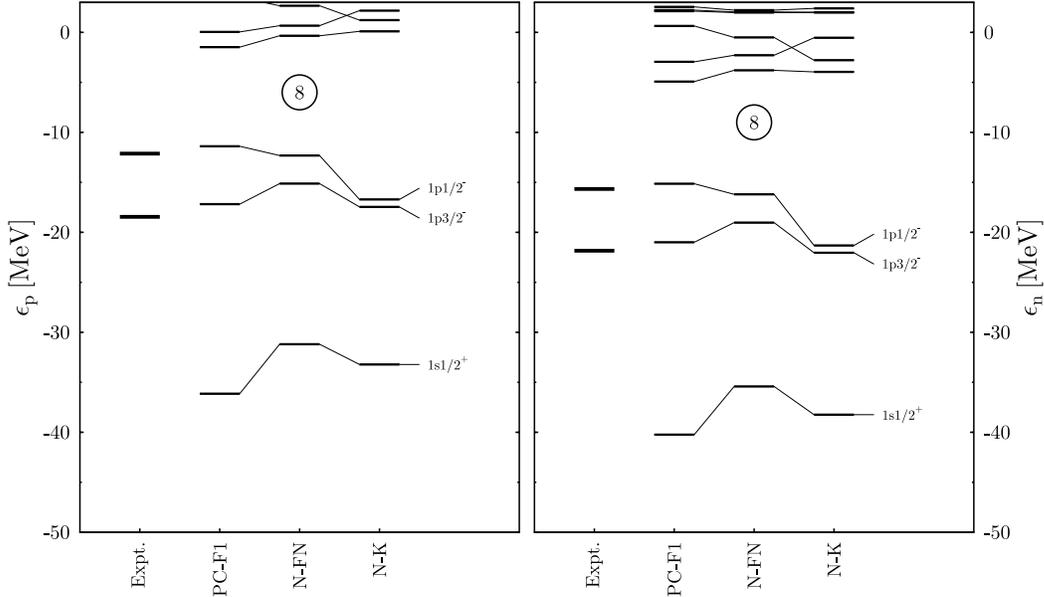,width=14cm}
\caption{Proton (left) and neutron (right) single-particle levels in $^{16}$O. On the left the experimental values for the p-states
are shown. Forces from left to right are PC-F1, N-FN, and N-K.}
\label{spec}
\end{figure}

Table \ref{obs} lists the experimental and predicted values of these observables for the sets
of coupling constants fitted
to finite nuclei as well as for the set determined by Koch et al \cite{Koch1}. Also, the total $\chi^2$
for the different sets is displayed.
\begin{table}
\caption{Sets of (rounded) coupling constants, cutoffs and explicit masses used in this study. The fourth
and fifth columns contain the naturalized coupling constants for two values of the
QCD large-mass scale $\Lambda$.}
\label{pars}
\begin{tabular}{l|cc||cc}
             &   N-K  &   N-FN  & N-FN($\Lambda=770$~MeV) &   N-FN($\Lambda=464$~MeV) \\\hline
	
 $g_1$    &   1.16 [fm$^2$]   &    1.775  [fm$^2$]    &    0.390   & 0.390  \\
 $g_2$    &   3.40 [fm$^2$]   &    3.118  [fm$^2$]    &    0.685   & 0.685\\
 $g_3$    &   0.92 [fm$^8$]   &    1.170  [fm$^8$]    &    0.189   & 0.066\\
 $g_4$    &   0.0  [fm$^8$]   &    0.471   [fm$^8$]   &    0.076   & 0.028 \\
 $g_5$    &   0.0   [fm$^8$]  &    2.787   [fm$^8$]   &    0.450   & 0.163  \\
 $g_6$    &   0.0  [fm$^4$]   &    -0.123   [fm$^4$]  &    -0.411  & -0.149   \\
 $g_7$    &   0.0  [fm$^4$]   &    -0.138  [fm$^4$]   &    -0.462  & -0.168  \\
 $g_8$    &   0.0  [fm$^2$]   &    0.562  [fm$^2$]    &    0.494   & 0.494  \\
 $g_9$    &   0.0 [fm$^4$]    &    0.167    [fm$^4$]  &    2.234   & 0.811   \\\hline
 $\it\Lambda_{\rm c}$  [fm$^{-1}$] &   3.25    &    2.35    & 2.35  & 2.35        \\
 $m_{0}$ [MeV] & 0.0     &    160.0     & 160.0 & 160.0    \\
\end{tabular}
\end{table}
We can see that the parametrizations N-FN constitute a major improvement over
N-K due to the additional terms and the inclusion of ground-state observables in the adjustment of the coupling constants. The original N-K model delivers radii and surface
thicknesses that
are too small. Due to the derivative and additional higher order terms, these observables are described better with
the new NNJL force N-FN.
The force N-K delivers a large amount of overbinding which is also cured in the new adjustment.
Nevertheless, N-K, which has been adjusted to nuclear matter properties only, is able
to qualitatively describe finite nuclei.
The average error with N-FN for
most observables lies in the range of a few percent or below. The surface thicknesses pose the greatest problem
while the nucleon mass, which is a prediction
and thus an additional demand on the model, is reproduced to within 0.24 \%.
 
We see, however, that the RMF-PC force PC-F1
is superior (its $\chi^2$ is smaller by a factor of 5 than that of N-FN) for the
ground-state observables. Pairing (which is included in PC-F1) would enhance the results from
N-FN, but not dramatically. The major deficencies of N-FN are the too large effective mass and compressibility.
 
The charge densities of the nuclei $^{16}$O, $^{48}$Ca  and $^{208}$Pb are shown in Fig. \ref{dens}.
The large density fluctuations of N-K have been cured in the new force, N-FN, though they are still too large in lead.
Also, the surface region that is much too small in a
model without derivative terms, like N-K, is predicted to be larger with N-FN, though relativistic models
still generally underestimate this observable \cite{Buer02}. PC-F1 comes closest to the experimental data, N-FN performs
second best.
 
Fig. \ref{spec} shows the proton and neutron single-particle levels of $^{16}$O for the forces
discussed as well as
the experimental values for the p states.
As has been inferred from the nuclear matter properties of N-K, the spin-orbit splittings with N-K
come out
way too small (the p-levels are almost degenerate), while they have the right size with PC-F1. N-FN
performs second best with the splitting being too small by approximately a factor of two. This relates
to the effective mass (0.75 vs. 0.61 for PC-F1) that is still a bit too large to deliver the
correct spin-orbit potential. Thus, the NNJL model displays deficiencies similar to the linear $\sigma$ models \cite{Fur93,Fur96}
The linear realization of chiral symmetry and its breaking by the NJL mechanism (as well as the prediction of
the nucleon mass) constitute strong constraints on the denstity dependence and size of the scalar and
vector potentials which seem to prohibit a degree of accuracy which has been reached with modern
relativistic mean-field models. Chiral models with a nonlinear realization \cite{Fur97,Sch02}, however, deliver an accuracy comparable to modern RMF approaches.
 
Table \ref{pars} shows the coupling constants and other parameters of the two models.
For the set N-FN we note that the isovector coupling constant $g_8$ has approximately the value of that of PC-F1,
which is $g_8 = 0.675$. This is due to the fact that $g_8$ is sensitive only to the difference of
neutron and proton vector densities and is thus unaffected by the size of the scalar density. The other 4-fermion coupling constants have different values due to the
fact that the scalar
density is much different from that of the RMF-PC approach given the inclusion of the condensate in terms
of nucleonic degrees of freedom.
 
As stated in the Introduction, QCD scaling provides a valuable test of the mean-field coupling constants and hence
of the compatibility of a model with low-momentum QCD. It has been applied successfully in nuclear physics (see \cite{Fri96} for the first application).
Naive dimensional analysis is used to obtain dimensionless coupling constants.
We use the scaling procedure of
Manohar and Georgi \cite{Mg84} but without pion fields (they vanish in the mean-field approximation), which reads
 
\begin{equation}
{\mathcal{L}} \sim -c_{l n}
\left[ \frac{\overline{\psi}\psi}{f^2_{\pi} \Lambda } \right]^l
\left[ \frac{\partial^{\mu}}{\Lambda } \right]^n
f^2_{\pi} \, \Lambda^2 \,
\label{IV.3}
\end{equation}
 
where $\mathcal{L}$ is a generic lagrangian term in the Weinberg expansion \cite{WE90}.
Here $\psi$ are nucleon fields
$f_{\pi}$ and $m_{\pi}$ are the pion decay constant, 92.5 MeV, and
pion mass, 139.6 MeV, respectively, $\Lambda= 770$ MeV is the QCD large-mass scale taken as the $\rho$
meson mass, and $\partial^\mu$
are usual derivatives.
Dirac matrices and
isospin operators (we use $\vec{t}$ here rather than $\vec{\tau}$)
have been ignored. Chiral symmetry demands \cite{We79}
 
\begin{equation}
\Delta = l + n - 2 \geq 0
\label{IV.2}
\end{equation}
 
\noindent such that the series contains only {\it positive} powers of
(1/$\Lambda $). A natural
lagrangian  \cite{Mg84} should lead to dimensionless coefficients $c_{ln}$ of
order unity.
Our more stringent definition \cite{Buer02} is that a set of QCD-scaled coupling constants
is {\it natural} if their absolute values are distributed about the
value 1 {\it and} the ratio of the absolute maximum value to the absolute
minimum value is {\it less than 10}.
Thus, all information on scales ultimately resides in the
$c_{lmn}$. If they are natural, QCD scaling works. In PC-F1, which has been obtained in an unconstrained optimization procedure with respect to nuclear ground-state observables, the 9 coupling constants are all natural
in terms of our stringent definition.
 
We have scaled the coupling constants of N-FN in two different ways. Firstly, we used the QCD mass scale of
$\Lambda = 770$~MeV (the $\rho$ meson mass) which has been used to scale the coupling constants of PC-F1.
Secondly, we scaled them with the scale $\Lambda = 464$~MeV$=2.35$~fm$^{-1}$ which corresponds to the
momentum cutoff of our model and thus constitutes a naturally occuring scale. The resulting scaled
coupling constants are shown in Table \ref{pars}. Note that the scaled 4-fermion coupling constants are
identical in both procedures since the mass scale does not explicitly enter in their scaling (this is not true for
the higher order terms). In both cases, the total set of coupling constants does  not fullfill
our criterion of naturalness. This is due to the ratio of the scaled values of $g_4$ and $g_9$ which exceeds 10.
Additionally, the absolute values do not scatter around $1$ but are overall smaller. The scaled
constants are approximately uniform and are, however, almost natural, that is, the deviations from 1 are approximately uniform and are not large.
We attribute their suppression to an incorrect magnitude and density dependence of the
potentials which also yields an effective mass that is too large. This result
shows that in the NNJL model QCD scales are not realized as well as in the established
RMF models, which is probably connected to the way chiral symmetry is both realized and broken here (RMF
 models
can be viewed as consistent with a nonlinear realization).
 
We note that with N-FN a fair description of nuclear ground-state observables and simultanesously the nucleon mass
occurs
even with a high value of the compressibilty (K=388 MeV). The nuclear ground-state observables do not appear to be very sensitive to its
value. The RMF-PC forces NL-Z2 \cite{Ben99} and NL3 \cite{Lal97}, which deliver good (and also comparable) results for finite nuclei have
quite different values, namely $K=172$ MeV for NL-Z2
and $K=260$ MeV for NL3. Calculations of excited states, however, favor the value of NL3.

\section{Conclusions}
We have presented an extended NJL model for both nuclear matter and finite nuclei. We tested
the compatibility of the ansatz and the NJL mechanism of mass generation due to spontaneous breaking of chiral
symmetry with nuclear bulk properties and ground-state observables of finite nuclei. We found compatibility in both cases.
 
In nuclear matter, we investigated the effect of a
small (160 MeV) explicit chiral symmetry breaking mass term as well as the cutoff dependence.
In least-squares adjustments, a cutoff of $\it\Lambda_{\rm c} \approx 2.0 - 2.75$~fm$^{-1}$
is favoured compared to higher values. It remains to be understood why the nucleon NJL model favours
such small values. Its similarity with the $\sigma$ meson mass value required in RMF-FR models deserves
further attention. An
explicit mass term of 160 MeV leads to better results compared to the chiral limit of the model. This corresponds to the
physically realistic case.
A too small cutoff and a large explicit mass term do not appear to be consistent with dynamically generated mass.
 
The results for finite nuclei show
that a simultaneous description of the nucleon mass and ground-state observables is possible:
reasonable values of $m_{0}$ and $\it\Lambda_{\rm c}$ could be found.
However, the overall quality of agreement with the measured observables
utilized in the least-squares adjustment is not as good as that obtained
in the RMF models.
 
An explicit mass term, the additional interaction terms, and the adjustment
procedure with respect to finite nuclei enhance the performance of the model to a great extent.
The fact that a not too large cutoff and $m_{0} = 160$~MeV are favorable can be understood from
the experience
with RMF-PC models: in the limit $m_{0} = 938.9$~MeV and $\it\Lambda_{\rm c} = 0$~fm$^{-1}$ we obtain the
RMF-PC model, apart from pairing,
with one exception: a 6 fermion term is not present due to the demands of chiral invariance
of the interaction terms of the NNJL lagrangian that we present.
The NNJL model in all cases tends to a rather high compressibility ($\approx 310-390$~MeV) in nuclear matter.
Though this value is too high, the description of finite nuclei is still fairly good.
The compressibility problem requires further study.
 
As expected, the model for finite nuclei with a small effective mass $m^*$ performs better than the model
with a larger effective mass: accurate spectral properties demand large scalar and vector
potentials and hence a small effective mass. Since the mean-field potential is only sensitive
to the sum of scalar and vector potentials, this situation could be changed with terms
that modify the spin-orbit potential, e.g. tensor terms.
For the sets adjusted with respect to finite nuclei, the violation of causality at high densities prohibits their application
far above saturation density, and we attribute this behavior to our phenomenological approach.
 
In its present form the
NNJL model performs reasonably, with the too large effective mass and compressibility perhaps being its major drawbacks.
However, there does not appear to be much room for improvement to remedy these drawbacks.
Obvious refinements include (a) a smooth 3-D or a covariant 4-D cutoff for the negative energy states, (b) pairing,  (c) extended
adjustment protocols, and (d) additional interaction terms (including different choices of $m_0$, for example the
pion-nucleon sigma term), but it is not clear that the size and density dependence
of the potentials will significantly change. In contrast to qNJL models, the applicability of this
model appears to be confined to nuclear systems in the mean-field approximation. Nucleonic
degrees of freedom and the current predicitve power of the NNJL model may not allow an extension to further areas of nonperturbative QCD.
 
An objective of this study was to test if the NJL mechanism of dynamic mass
generation due to spontaneous symmetry breaking of linearly realized chiral symmetry translates, in some way, from quarks to nucleons.
Our correspnding NNJL model has not failed in this endeavor, but instead has had
limited success. This suggests that pursuit of the linkage between low-momentum QCD
and the nuclear many-body problem should be continued. However, with regard to
the description of nuclear ground-state properties and the use of QCD scales, the nonlinear realization of chiral symmetry
is superior compared to linear models like the present NNJL approach or the existing $\sigma-\omega$ models.

\section*{Acknowledgements}
The authors thank V. Koch and the referees for useful and helpful comments. This work was supported by the U.S. Department of
Energy.

\end{document}